# Impact of Smartphone Distraction on Pedestrians' Crossing Behaviour: An Application of Head-Mounted Immersive Virtual Reality

**Anae Sobhani*** (Corresponding author), Postdoctoral Fellow
Transport and Logistics group, Faculty of Technology, Policy and Management, Delft University of Technology
Jaffalan 5, Building 31, Delft, 2628 BX, Netherlands
Ph: (0)15 27 88384
Email: anae.sobhani@mail.mcgill.ca

**Bilal Farooq,** Assistant Professor
Laboratory of Innovations in Transportation (LITrans), Department of Civil Engineering, Ryerson University
Canada Research Chair, Disruptive Transportation Technologies and Services
341 Church Street, Toronto (ON), Canada, M5B 2M2
Ph: 416-979-5000 Ext: 6456
Email: bilal.farooq@ryerson.ca

**ABSTRACT**

A novel head-mounted virtual immersive/interactive reality environment (VIRE) is utilized to evaluate the behaviour of participants in three pedestrian road crossing conditions while 1) not distracted, 2) distracted with a smartphone, and 3) distracted with a smartphone with a virtually implemented safety measure on the road. Forty-two volunteers participated in our research who completed thirty successful (complete crossing) trials in blocks of ten trials for each crossing condition. For the two distracted conditions, pedestrians are engaged in a maze-solving game on a virtual smartphone, while at the same time checking the traffic for a safe crossing gap. For the proposed safety measure, smart flashing and color changing LED lights are simulated on the crosswalk to warn the distracted pedestrian who initiates crossing. Surrogate safety measures as well as speed information and distraction attributes such as direction and orientation of participant's head were collected and evaluated by employing a Multinomial Logit (MNL) model. Results from the model indicate that females have more dangerous crossing behaviour especially in distracted conditions; however, the smart LED treatment reduces this negative impact. Moreover, the number of times and the percentage of duration the head was facing the smartphone during a trial and a waiting time respectively increase the possibility of unsafe crossings; though, the proposed treatment reduces the safety crossing rate. Hence, our study shows that the smart LED light safety treatment indeed improves the safety of distracted pedestrians and enhances the successful crossing rate.

# 1 INTRODUCTION

Although walking is perceived as an automatic thought process, it in fact entails attention to surrounding environmental information and details to act upon. Past studies show that adult's performance and efficiency decrease when multi-tasking especially for seniors (see (Kramer & Madden, 2008)). Moreover, studies on different adult age categories in this context indicated that dual-task decreases the memory encoding and walking speed in general for everyone but more considerably for seniors and mid-aged adults (Mark B Neider et al., 2011). That said, in the past decade many studies have focused on the effects of using cell phones on road users' reactions in the transportation planning field. Findings show that drivers are significantly impaired when using their cell phone. Also, the probability of a pedestrian crossing the intersection successfully decreases, which in some cases leads to unsafe crossing attempts. These studies make use of either real observed or simulated environment data for their analysis.

In the field of transportation studies, pedestrian safety has been a topic of interest for many years, however, despite the vast number of studies and implemented safety measures, pedestrian road injuries and fatalities are still among the top causes of death. The most recent World Health Organization study on road traffic crashes reported that globally, vulnerable road users suffered roughly 625,000 fatalities and 25 million injuries (World Health Organization, 2016). Canadian statistics from 2005 reported 11.8% of all road fatalities and 10.4% of all road injuries to be pedestrians (Ifedi, 2005). Comparing these values to the following decade, we see that in 2015, pedestrian fatality increased to 15.2% of all road casualties, and pedestrian serious injury had an increase to 14.3% of total road injuries (Transport Canada, 2015). One of the controversial reasons for this increase is the rise in distracted pedestrians, who are on their phones talking, texting, surfing the web, looking for directions, or playing games (J. Nasar, Hecht, & Wener, 2008; J. L. Nasar & Troyer, 2013; Tapiro, Oron-Gilad, & Parmet, 2016). Before the widespread use of smartphones and other portable technologies, pedestrian violations were among the top injury and fatality contributors (Transport Canada, 2010). In recent years, violations such as J-walking, crossing during the red-light phase and failure to yield to vehicles is combined with distractions related to smartphone use which creates a high-risk situation resulting in an increase in pedestrian road injury and fatality. Hence, some studies have focused on different types of pedestrian crossing facilities to address the unsafe crossing concerns (e.g. see (Anciaes & Jones, 2018; Gitelman, Carmel, Pesahov, & Hakkert, 2017)). However, this increase in fatality and injury rate is alarming given the numerous studies, policies, educational and safety measures, improvements and implementations aiming to reduce these risks.

Unsignalized road crossing, although thought of as a simple task, demands an individual's undivided attention and concentration. To successfully select a gap that allows for safe crossing, individuals must accurately judge the spatial and temporal size of the gap in relation to their own ability to cross the gap in time. This process is further complicated when there is more than one lane of traffic (Plumert & Kearney, 2014). When traffic flow increases in both directions, safe crossing gaps become smaller and less frequent (T. Wang, Cardone, Corradi, Torresani, & Campbell, 2012). Since crossing violations are in part due to added pressure from temporal restraints, the act of crossing is usually rushed, resulting in more careless crossings. When the cognitive demand required for crossing is divided by distractions, the pedestrian's awareness is reduced resulting in unsafe and risky crossing behaviours (Banducci et al., 2016; Lin & Huang, 2017; J. Nasar et al., 2008). As a result of the increase of distracted pedestrian, different counter measures have been proposed such as the development of smartphone applications that display warning signs on the phone of the distracted pedestrian when initiating a road crossing.

To study the behavioural and safety aspects of distracted and undistracted street crossings, several methods have been used. The use of visual attributes in transportation safety studies have caused controversy in literatures (see (Farooq, Cherchi, & Sobhani, 2018; Olshannikova, Ometov, Koucheryavy, & Olsson, 2015)). To study perception, past visualization tools relied on pictures, photomontages, maps or simulation videos (e.g. see (Song, Lehsing, Fuest, & Bengler, 2018). An emerging interactive technology in the field of transportation studies are Virtual Reality (VR) tools which have opened a new window for practical applications and scientific investigations of human perception and behaviour. VR allows the user to immerse in a controlled environment for in-depth evaluation of user perception and behaviour. Its ability to overlay the physical environment with virtual elements such as information or images, and to allow participants to interact with the physical environment in real time, provides new possibilities for content delivery. VR allows the sensation of immersion in the activities on the screen and the virtual elements (Animesh, Pinsonneault, Yang, & Oh, 2011; Faiola, Newlon, Pfaff, & Smyslova, 2012; Farooq et al., 2018; Jennett et al., 2008; Nah, Eschenbrenner, & DeWester, 2011). Moreover, recent advances in computer graphics and technology have provided new opportunities for generating more realistic virtual scenarios that are suitable for behavioural studies (Patterson, Darbani, Rezaei, Zacharias, & Yazdizadeh, 2017). VR environment experiments have successfully been conducted in various fields of cognitive studies (Blissing, Bruzelius, & Eriksson, 2017; Farooq et al., 2018). For example, Lehsing and Feldstein analysed social interaction in transient where a car driver in a driving simulator encounters a pedestrian in a second simulator in varying situations by adapting VR as their survey tool (Lehsing &

Feldstein, 2018). Also, studies have shown that people can develop realistic spatial knowledge in the VR environment that is similar to the actual physical environment (O'Neill, 1992; Perroud, Régnier, Kemeny, & Mérienne, 2017; Ruddle, Payne, & Jones, 1997; Rusch et al., 2013; Tlauka & Wilson, 1996). The main advantage of adapting the VR in research studies is the freedom and versatility in setting up experiments which enables scientists to measure not only perception, but physical reactions of participants by adopting electrocardiography, skin conductance, electroencephalogy, and eye-tracking (Patterson et al., 2017; Wiener, Hölscher, Büchner, & Konieczny, 2012). A Head-Mounted VR display device uses an optical system to directly present virtual scenes received by the display and works with the human brain to produce a strong sense of immersion (Y. Wang et al., 2016). Feldstein et al. used a combination of a driving simulator along with a visual based motion capture system with a head-mounted virtual reality device to link a driving simulator to a pedestrian simulator so that both participants meet in the same virtual environment to analyse their interactive behaviour simultaneously (Feldstein, Lehsing, Dietrich, & Bengler, 2016). Immersive virtual environments which allow a locomotive interface may preserve the perception-action coupling that is critical in examining many visual timing skills (Wu, Ashmead, & Bodenheimer, 2009). Researchers also applied virtual environments to study pedestrian route choice and reaction to information in evacuation scenarios. However, most of these studies lack the extensive two-way interactivity and dynamics necessarily to create realistic experiments. Specifically, there is a strong need for the incorporation of user's actions within the environment and dynamic responses by various elements of the environment to user's actions e.g. if a user walks in front of a virtual car, it should change direction or slowdown in accordance with the expected behaviour.

In this paper, to better understand the behaviour of pedestrians looking at their phones when crossing a road, we developed an experiment within the novel Virtual Immersive and interactive Reality Environment (VIRE) developed in our lab for a range of transportation experiments. Simulation of virtual vehicles, pedestrians, cyclists, and other objects in VIRE is dynamic, interactive, and ensures that virtual objects react to user's actions. Furthermore, VIRE has the ability to replicate personal object e.g. smartphone in hand and its usage (see (Farooq et al., 2018) for details). Major benefits of using virtual environment is the ability to examine various scenarios for incidents, employ appropriate treatments/policies and analyse their performances before making irreversible decisions that are either difficult or expensive in the real world. Three scenarios are proposed in order to perform our safety analysis and preventative measure evaluation: 1) pedestrians cross the road with no distraction, 2) pedestrian cross the road while solving a maze-puzzle on a smartphone, and 3) pedestrians cross the road while solving a maze-puzzle on a smartphone with smart flashing and color changing LED lights (white changes to blue when distracted crossing is initiated) installed on a cross road as a counter measure.

The following section (Section 2) summarizes the literature relevant to the paper and the current study in context. Section 3 describes the methodology behind the study which covers the general description of the virtual environment simulation system design, and the model structure. In Section 4, the variables used in our analysis are presented and a brief descriptive and sensitivity analysis of the collected data in VIRE is discussed. Section 5 covers the discussion of our findings across the three crossing conditions. Finally, the work is summarized in Section 6 and the conclusion, limitations and potential future studies are presented.

## 2    BACKGROUND AND STUDY CONTEXT

### 2.1 Past Studies

Going over the literature, there are four features of interest in safety studies of distracted road users: 1) type of distraction, 2) quality of the virtual environment, 3) safety evaluation tool, and 4) preventative measures. As the first feature of interest, types of distractions include looking at phone, conversing on the phone or listening to music. The cognitive demand related to talking, texting, surfing the web and social media as well as playing games on smartphones limits the attention from analysing scenarios and making appropriate decisions (Banducci et al., 2016; Yager, Cooper, & Chrysler, 2012). Many studies have investigated the effects of distracted pedestrian and drivers on their decision-making process (see e.g. (M. Cunningham and Regan 2018; M. Cunningham and Regan 2017; Haque and Washington 2015; Oviedo-Trespalacios et al. 2016)). Studies made use of simulated road crossing tasks while conversing on a smartphone as well as texting and driving (see (Gaspar et al., 2013; Mark B Neider et al., 2011)). All of which have shown an increase in crash likelihood when either the driver or pedestrian is distracted.

There have been several approaches to evaluating the safety of distracted pedestrian's road crossing behaviour. Some studies employed observational methods and manual data collection of real time pedestrian crossings (Banducci et al., 2016; Hacohen, Shvalb, & Shoval, 2018; Hatfield & Murphy, 2007; J. L. Nasar & Troyer, 2013). In addition, there have been many studies (Banducci et al., 2016; Brisan, Vasiu, & Munteanu, 2013; Dommes, Cavallo, Dubuisson, Tournier, & Vienne, 2014; Farooq et al., 2018; Ye, Xiao, & Yang, 2017) that made use of virtual environments due to their attractive features of being controlled for specific design scenarios and pose no risk to participants. Muehlbacher et al. studied the benefits and challenges of multi-road user VR simulations compared to single driver simulations, and at the end recommended the case where the multi road user simulation was appropriate (Muehlbacher, Preuk, Lehsing, Will, & Dotzauer, 2018). Although laboratory studies may not be completely representative of the real world, it is the closest environment where participants can relate to without being positioned in an unsafe real world situation, and therefore, they provide the most reliable alternative for these studies (Holland & Hill, 2010).

Hatfield and Murphy conducted an observational study of pedestrian crossing behaviour while using a smartphone compared to not using one (Hatfield & Murphy, 2007). This study identified risky behaviour from the crossing pedestrian, such as whether the pedestrian looked for traffic before starting to cross, whether there was an aggressive behaviour from either pedestrian or driver such as running to avoid the crash, or severe breaking of the driver, where the pedestrian looked when crossing, etc. as the potential factor that would influence safety (Hatfield & Murphy, 2007). Their results suggest that talking on a mobile phone is associated with cognitive distraction that may undermine pedestrian safety (Hatfield & Murphy, 2007). Nasar and Troyer used accident data involving a pedestrian distracted by their smartphone and found an increase in pedestrian injury over 6 years due to the increase in smartphone distractions (J. Nasar et al., 2008; J. L. Nasar & Troyer, 2013). Another study evaluated crossing behaviour of different age groups conversing on their smartphone while crossing the road in a semi-immersive virtual environment where pedestrians had to push a button when they felt it was safe to cross (Tapiro et al., 2016). The virtual environment was set up as a 180° spherical screen with three projectors casting the road environment (Tapiro et al., 2016). Their safety analysis was based on visual attention distribution using an eye tracking device fixed to the participant's forehead, participant's response time, Post-Encroachment-Time (PET) etc. were collected (Tapiro et al., 2016). Results based on a general linear mixed model showed that although different age groups showed different behaviours, all age groups were at a higher risk when conversing on their phones at the time of crossing (Tapiro et al., 2016). Byington and Schwebel studied a group of young adults who were replying to emails on their phone while crossing a road in a virtual environment consisting of three large monitors, arranged in a semi-circle providing a 180◦ field-of-view on the two-lane road (Byington & Schwebel, 2013). Focusing on safety measures such as start delay, wait time, missed opportunities, looked at traffic, and eyes off road, they found that while distracted participants took longer to cross, they chose riskier times to cross and missed more safe crossing opportunities (Byington & Schwebel, 2013). Schwebel et al. made use of an interactive, semi-immersive virtual environment for four types of road crossings: pedestrian conversing on their phone, pedestrian crossing while texting, and pedestrian crossing while listening to music, and pedestrian crossing while not distracted (Schwebel et al., 2012). Safety measures such as looking at traffic, looking away, hits, missed opportunities, and PET were utilized, and their findings showed that participants who were distracted by music or texting were more likely to be hit by a vehicle compared to undistracted participants (Schwebel et al., 2012). Moreover, using a high-fidelity street crossing simulator, Neider et al. showed that naturalistic smartphone conversations impair crossing performance and increase crash rates (M. B. Neider, McCarley, Crowell, Kaczmarski, & Kramer, 2010). Similarly, Stavrinos et al. demonstrated significant costs to simulated crossing performance while conversing on a smartphone (Stavrinos, Byington, & Schwebel, 2009, 2011). An innovative approach by Zaki and Sayed made use of video data and computer vision techniques to perform a preliminary study of automatically detecting distracted pedestrians on crosswalks using their gait parameters which eliminates the costly method of manually studying pedestrians on real crossing situations (Zaki & Sayed, 2016).

Several different preventative measures have been proposed, for example, the development of Safety-Aware Navigation Application (SANA) based on road user GPS data, which gives warning to both vehicles and pedestrians as soon as they arrive in the range of the safety distance (Hwang, Jeong, & Lee, 2014). Another mobile application was developed and tested in a virtual environment where an alert is displayed on the distracted pedestrian's smartphone screen who is about to enter a potential danger zone when crossing the road (Chen, Zhu, & Wang, 2012). The CrowdWatch mobile application also displays safety alerts based on the pedestrian's acceleration, orientation, GPS data and is in sync with traffic light and dynamic barriers to alert pedestrians about potential dangers (Q. Wang, Guo, Peng, Zhou, & Yu, 2016). A study by Hakkert, looked at the evaluation of crosswalk warning systems and their effects on pedestrian and vehicle behaviour (Hakkert, Gitelman, & Ben-Shabat, 2002). Another approach was employed using the back camera of a smartphone to detect vehicles approaching the user while they are conversing on their phone (T. Wang et al., 2012). A study by Zhou computed walking pattern recognition system and applied it to a smartphone application HeadsUp to warn pedestrians and locks the screen when the pedestrian is looking at their smartphone while walking (Zhou, 2015).

**2.2 Current Research Study in Context**

Since past studies have identified interacting with the phone (texting, emailing, reading or looking for directions) as the riskiest form of distraction, our study will make use of a maze-solving puzzle in a smartphone as the distraction source. Focusing on the second feature mentioned before, the closer the virtual environment is to reality and the more immersive it is, the more realistic participants' choices will be. For our virtual environment, we utilized VIRE where the participant can have a fully immersive and interactive experience at a road crossing that has been built as a replicate of a real Montréal urban road. Using this environment also provides us with the participants head and smartphone angle enabling us to study the number of times the pedestrian is looking up to evaluate the traffic situation or looking down at the phone. In our study, as the third feature, we employ surrogate safety measures. Since not all risky behaviour will result in a crash, surrogate safety measures are good indicators for identifying the risk level of the crossing decision, such as post-encroachment time (PET), time-to-collision (TTC), as well as acceleration and deceleration rates. The preventative measure features in most studies involve a mobile application and although many of these applications are available to download, the majority of the population might not install them. In addition, most of these applications are preventative given only a single distraction such as texting or conversing on the phone. Our approach makes use of smart flashing and color changing LED lights installed at road crossings to alert any pedestrian who is distracted by their phones, which is visible to them whether they are looking at their smartphone, talking on their phones or listening to music. This trend of safety counter measures has been used recently in several cities worldwide (e.g. Vancouver, Shanghai, Bodegraven-Reeuwijk, etc.). It should be noted that in this study we only focus primarily on the interactions between vehicles and pedestrians exclusively.

To the best of our knowledge, the combination of fully immersive and interactive virtual environment (i.e. VIRE) with three crossing conditions including a safety treatment investigating the influence of smartphone distractions on behaviour and surrogate safety measures has not been investigated before.

## 3  METHODOLOGY

In this section we first describe the development of the experiment within VIRE, after which the discrete choice model specifications used for the empirical analysis are presented.

### 3.1 Simulation System Design

In this section the participants' characteristics, virtual environment, traffic simulation, control rules, and street crossing paradigms in our study are described in detail.

#### 3.1.1  Virtual Environment

The unsignalized intersection environment used here is a simulation of an existing crossing in Montréal with high pedestrian activity (Sobhani, Farooq, & Zhong, 2017) (see Figure 1). The chosen road crossing location has high pedestrian activity levels because of the Laurier metro station and the presence of many shops, restaurants and cafés. The road crossing area is a one-way single lane road which is 5 meters long and has bike lanes on either side of the road. Furthermore, avenue Laurier is one of the accessible bicycle and pedestrian environment in the city with designed green and wide sidewalks.

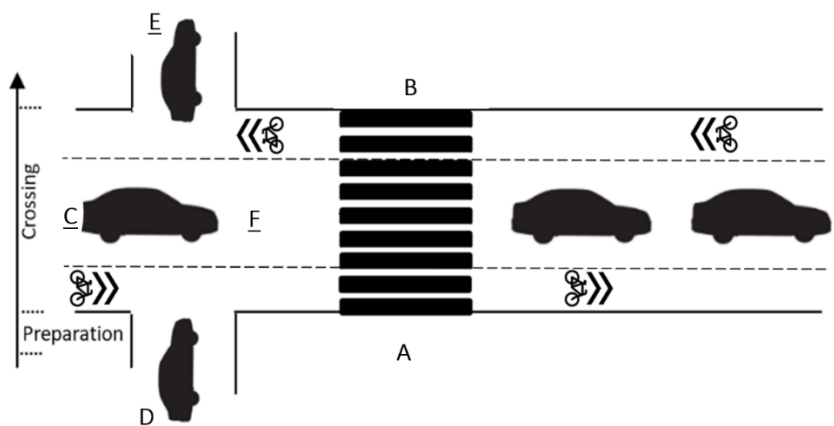

**Figure 1: The Road Street Crossing Structure**

Since the aim of this paper is to quantify the interactions between vehicles and pedestrians exclusively, we simplify the simulation to remove other pedestrians and cyclists. The crossing in our experiment is from point B to A, which is a distance of 5 meters and 2.5 meters wide. Pedestrians cross between B and A, while vehicles move through points

C - F, as well as D - E. Right turning vehicles move from point D to F, and left turning vehicles drive through points C and E. In reality, pedestrians would assess the traffic flow and behaviour for these four movements to identify a safe crossing situation. In our simulated environment, the traffic pattern simplified at this stage where vehicles are coming from the one entry leg (C). Furthermore, as being a side residential street, the traffic volume from Rivard street is negligibly small.

As mentioned, this paper adopts VIRE with the 3D architecture and scale built to represent the described intersection in order to provide a controlled simulation setting that is not possible in the actual physical location. The VIRE device is presented in Figure 2 and 3 showing the Oculus rift (headset and earphones), the Oculus touch controller (shown as smartphone in the VIRE environment) in the participant's hand used to solve the maze-puzzle, and the sensors. For more realistic effects, an audio of the moving vehicle was integrated in our system with respect to the distance and speed of the moving vehicles.

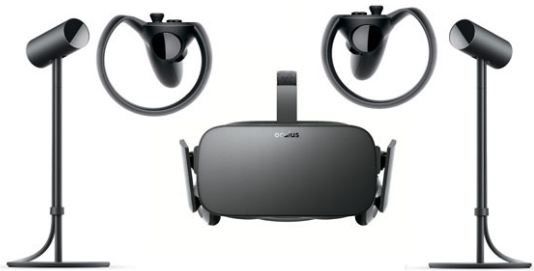

**Figure 2: VIRE Gear (Rift, Touch controllers, and Sensors**) ("Oculus Gear VR," n.d.)

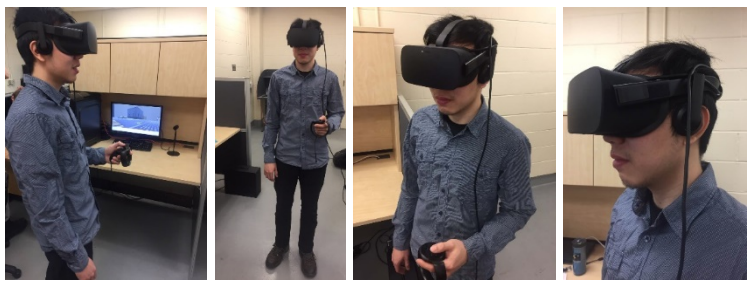

**Figure 3: VIRE Demonstration**

The simulation is separated into two main parts. The first part uses the unity engine for visualization in virtual reality and its physics engine. The unity engine supports stereoscopic virtual reality on Oculus Rift. The implementation is designed to detect the orientation and position of the user's head and hands via a set of sensors. Every agent is separated into a physics based controller and a facade to the agent in the actual model. The information it takes from the model for any car is the acceleration and orientation of the steering wheel, from which the speed, position and orientation of the vehicle is calculated and recorded in the simulation. Various 3D models and textures has been placed in the simulation for realism purposes.

The other part of the simulation is an implementation of the road environment and agent based simulation using various models. At every iteration of the simulation, which is done approximately every 60th of a second, unity takes the output of the model, applies it via its physics engine, and feeds the results back to the model.

The simulation was coded in Java and C++, and for visualization the unity gaming engine was adopted. For implementation of the experiments, a computer with Core I7 6700K (4.00GHz 8M Cache) and Nvidia GeForce GTX 1080 along with Oculus Rift, motion and touch sensors was used.

3.1.2    Traffic Simulation

The traffic simulation system is an agent based multimodal traffic microsimulation which interacts with the VIRE and responds to user input and actions (tracking motions). The vehicular traffic is simulated using car-following model, while virtual pedestrians are simulated using social force model. Both models have already been calibrated using real datasets.

Since the system is agent based, the gaps between cars and their path were predetermined with a Poisson distribution (μ=4 s) and also two safe gaps (5 and 7 seconds) based on past literature were randomly forced in the simulation system to give participants at least two safe crossing opportunities. It should be noted that the measured gap between any two cars was generated as the difference between the time that the front bumpers of the first vehicle and the vehicle behind it pass a certain point. Based on Ville Montréal's traffic database, the maximum approach speed of vehicles at this intersection is 50 km/h, and the traffic volume during peak hours at the interaction (only straight C – F path (see Figure 1) is 1080 vehicle per hours. We simulate a relatively high vehicle flow where cars can speed up to the mentioned maximum speed. Thus, simulated vehicles gradually accelerate from the approach speed (0 km/h) to the maximum speed and then maintain it. Further, the vehicle-to-vehicle and vehicle-to-pedestrian (participants) interactions had to be defined carefully (see Figure 4).

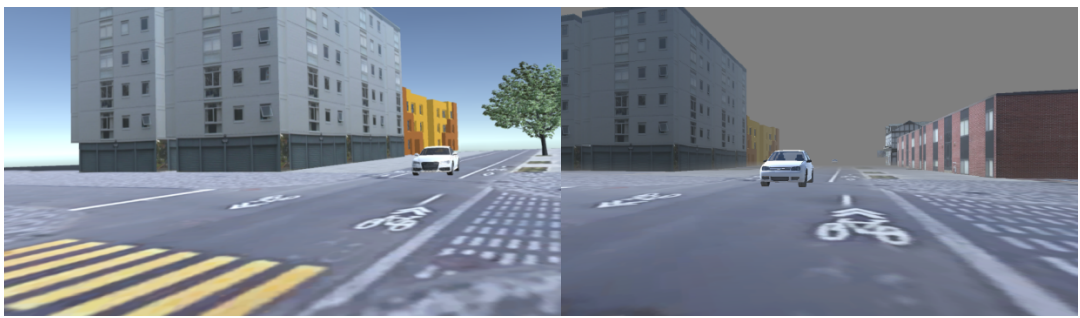

**Figure 4: Simulated Road Crossing from the Perspective of the Pedestrian Looking at Traffic**

Thus, as for the simulation, the network was drawn from the OpenStreetMap data and the interaction between vehicles was done with a simple car following model. At every iteration, the behaviour of a car was based on its desired goal, its speed, position and orientation, along with the speed, position and orientation of the car in front of it. In each trial, every car checked every object in its local vicinity to react appropriately depending on this agent's characteristics. In other words, to avoid collisions, vehicles accelerated or slowed down so that they stayed in the predetermined sight distance to the foregoing vehicle. Like reality, vehicles entering the simulated road yield to cars or pedestrians on the street or crosswalk area to maintain the safe distance. It should be noted that geometric characteristics of the road also was considered in the modeling movement of vehicles.

3.1.3    Street-Crossing Paradigm and Experiment

The experiment session was divided into two sub-sessions (i.e. the training and the actual session) which took up to one hour (see Figure 5).

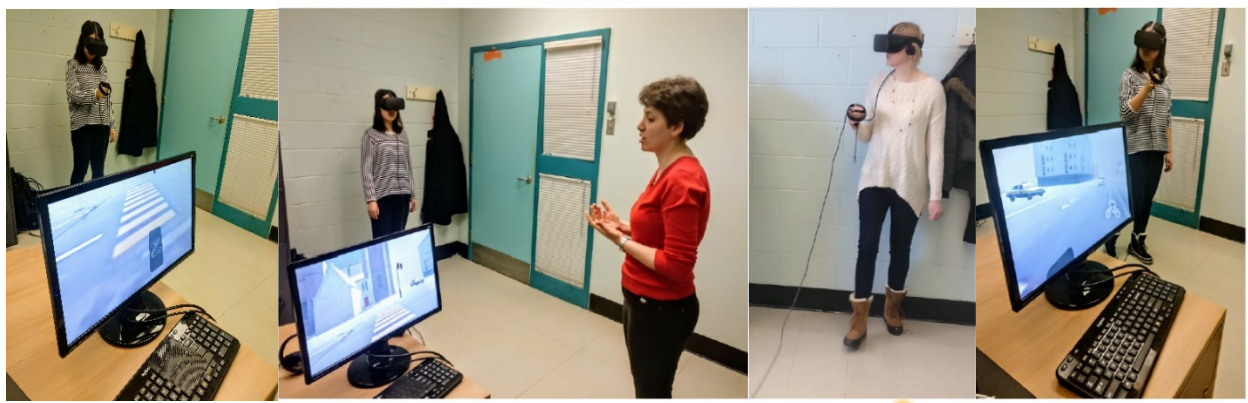

**Figure 5: Respondent in VIRE**

The training session was held to familiarize the participants with the simulated intersection in the VIRE. Also, in this session, they were trained on the tasks in a three-step process. First, they were asked to use the Oculus touch controller to solve a simple maze on the smartphone application interface (see Figure 5). Then, they came to know the virtual intersection environment by looking around and traversing the crosswalk back and forth without traffic. Then an example of traffic flow was shown to them. Finally, each participant completed six practice trials of the street crossing tasks during which they were informed that despite the fact that cars would decelerate for them during crossing the street, they were not allowed to make the car do so. In other words, they were responsible for seeking a safe and comfortable gap to cross for themselves and the cars. For the purpose of our study, running to cross the road was not allowed.

In the actual experiment session, participants had to perform 30 successful crossing trials, 1) 10 successful crossings with no distractions, 2) 10 successful crossings while participants had to simultaneously solve as many maze-puzzles as fast as they could, and 3) 10 successful trials with the implemented the smart LED safety measures while participants were distracted with the maze-puzzle. In each trial, participants were initially directed to the crossing point where the simulated traffic stream would pass by. Participants were allowed to rest between trials. The experiment continued until 10 successful crossings were recorded for each condition. In other words, in each trial, participants stood at the edge of the sidewalk of the busy street (avenue Laurier) and observed the traffic approaching from their right side (B to A in Figure 1), and finally crossed when deemed safe. Each trial was a maximum of 60 seconds and ended whenever the participant finished crossing. If the participant did not cross during the one minute of the trial, it was recorded as a time-out trial. Also, whether the participant got into an accident or made the vehicles in the road yield to him/her significantly, it was recorded as a failed trial. In the occurrence of these events, the same scenarios eventually were presented to the participant until he/she made a successful crossing. It should be noted that volunteers were not visually or audibly informed regarding crossing success or failure at any time.

### 3.1.4 Participants Description of Participants

The participants were recruited from Montréal and Toronto from February to the end of May 2017. Forty-two volunteers participated in our study from Montréal and Toronto in total. A written consent before the experiment was provided to the subjects, and the procedure was approved by the Institutional Ethic Board of both universities. Moreover, a questionnaire was designed to obtain general information such as the likelihood of the subjects' engagement with their smartphone while crossing a street.

Out of the forty-two participants, 59.9% are from Montréal, 40.5% are females, and the age range is between 18 to 45 years. They all had normal or corrected to normal binocular visual acuity and hearing. While 62% of them never used VR before, the rest had no significant amount of experience with virtual environments (i.e. 70% used VR less than 4 times). Road crossing walking speeds were recorded by the VR which indicated that the average walking speed was 1.1 m/s. This observed attribute is in accordance with several field studies (see e.g. (Dommes et al., 2014; Knoblauch, Pietrucha, & Nitzburg, 1996)).

### 3.2 Multinomial Model Structure

The empirical framework for our study employed the Multinomial Logit model (MNL) for each crossing condition. Let us consider $k = (1, 2, \ldots, K)$ as a decision maker, $J$ as the number of alternatives, and the following functional form for utility ($U_k(\boldsymbol{x})$):

$$U_{kj} = \beta x_{kj} + \varepsilon_{kj} \tag{1}$$

where $x_{kj}$ is the observed attribute affecting the utility function of the alternative choice of each individual, $\beta$ is the column vector of coefficients in associated with $x_{kj}$, and ($\varepsilon_{kj}$) is called the random term which characterizes the idiosyncratic effect of unobserved variables. This term is assumed to be independent of $x_{kj}$.

The probability of individual $k$'s observed choices, conditional on $\beta$ is (Mcdowell & Shi, 2014):

$$P_{kj} \mid (\beta) = \frac{e^{\beta x_{kj}}}{\sum_{j=1}^{I} e^{\beta x_{kj}}} \tag{2}$$

### 4 DATA AND DESCRIPTIVE ANALYSIS

This section presents the study variables, collected data and descriptive analysis of the three study conditions.

## 4.1 Variables

For analyse purposes, based on past literature, four types of variables were generated from the collected crossing behaviour data: 1) crossing variables: crossing duration (time from initiating the crossing until the crossing was complete), wait time duration (the time from the start of the trial where the participant was standing on the sidewalk until the crossing was initiated), crossing speed, and initiate crossing speed, 2) distraction attributes: percentage of the time the head was facing the smartphone during wait time, percentage of the time the head was facing the smartphone while crossing, number of times the head was facing the smartphone over the total trial duration, 3) safety measures: TTC (calculated for each second between each vehicle and the pedestrian as the time until they would collide if their direction and speed remain unchanged), PET (the time difference between when the pedestrian departs the collision point and the vehicle arrives at that point), maximum acceleration, maximum deceleration, and the percentage of successful, failed or time-out trials, and 4) socio-demographic information: age, gender, number of years they owned a smartphone, having a driving license. It should be noted that for modeling proposes, in addition to estimating a variable's impact on PET (depend variable), the deviations of reasonable variables were evaluated and estimated through interaction variables, for example Female * Maximum acceleration, Female * Percentage of the time the head was oriented toward smartphone during crossing, etc. The model specification was arrived at through a systematic process of removing statistically insignificant variables and combining variables when their effects were not significantly different.

To analyse the PET and TTC factors, dangerous conflicts are defined as interactions with TTC or PET below 1.5 seconds. This threshold has been defined arbitrarily and an extensive discussion regarding its range can be found in (Zaki & Sayed, 2016). It is worth mentioning that other threshold values have been examined and the results were found to be robust.

In the following section the interaction between these variables are studied to discuss the behaviour pattern across the three different crossing conditions.

## 4.2 Descriptive Analysis

Table 1 presents the mean value of the generated crossing attributes across different conditions. Looking at the data for the not distracted condition (condition one) we observe that pedestrians waited on average 18.0 seconds (st. dev. = 9.6s) on the sidewalk before initiating crossing. The distracted pedestrian condition (condition two) and the distracted pedestrian with smart LED safety measure (condition three) waited on average 21.2 seconds (st. dev. = 10.9s) and 21.3 seconds (st. dev. = 11s) respectively. These large standard deviations show that in our data, there are participants who were more patient versus some participants who did not need more time to cross. That said, in general, distracted participants waited 18-19% longer to cross compared to not distracted. This is reasonable since distracted pedestrians were engaged with their phones while also checking the traffic for a safe crossing gap. Comparing the distracted pedestrian where there are smart LED lights on the crosswalk, do not show a significant difference (1%) regarding the waiting time behaviour compared to condition two (distracted with no safety treatment). Looking at the behaviour between females and males, it is observed that non-distracted females waited on average 20.9 seconds while males waited on average 16 seconds. Comparing the pattern among the conditions indicates that while distracted, both genders waited longer to initiate crossing compared to not distracted condition. In addition, comparing the average wait time based on gender illustrates that females in condition three waited more than condition two while men's behaviour did not change that much. Looking at the age component, in all three conditions, older participants (30 years and older) took more time before crossing compared to younger adults (between 18 and 30 years old). This indicates that younger pedestrians are less patient to cross.

Another interesting behaviour changes is observed in the successful crossing time duration and average walking speed. In general, it took less time to successfully cross the street when pedestrians were distracted (crossing duration of 4.4 seconds, 4.2 seconds and 3.8 seconds for condition one, two and three respectively) (see Table 1). This is due to not looking at the road while walking distracted, so they increased their speed to complete their crossing. Females did not have a significant change in speed across different cases, while male participants walked faster when they were distracted (condition two and three).

Moreover, analysing the average initiating walking speed indicates that both genders did not have significant changes in initial speeds across different cases. However, females generally had a lower speed when initiating the crossing compared to males.

The second set of variables that was studied in this paper was the distraction attributes. In general, it was observed that in the case with the smart LED flashing lights, people spent more time on their phone while waiting at the sidewalk (74.4%) but were more aware of the environment around them (looking at the road) while crossing (69.6%) compared to untreated distracted condition (see Table 1).

| | Variables | | Condition | General | Female | Male |
|---|---|---|---|---|---|---|
| **Crossing** | **Wait time duration** *(s)* | | Not distracted | 18.0 | 20.9 | 16.0 |
| | | | Smartphone | 21.2 | 23.9 | 19.4 |
| | | | Distracted & safety measure | 21.3 | 24.4 | 19.2 |
| | **Crossing duration** *(s)* | | Not distracted | 4.4 | 3.9 | 4.7 |
| | | | Smartphone | 4.2 | 4.1 | 4.3 |
| | | | Distracted & safety measure | 3.8 | 3.7 | 3.8 |
| | **Crossing speed** *(m/s)* | | Not distracted | 1.0 | 1.0 | 1.0 |
| | | | Smartphone | 0.9 | 1.0 | 0.9 |
| | | | Distracted & safety measure | 1.0 | 1.0 | 1.1 |
| | **Initial walking speed** *(m/s)* | | Not distracted | 1.6 | 1.5 | 1.6 |
| | | | Smartphone | 1.5 | 1.4 | 1.6 |
| | | | Distracted & safety measure | 1.6 | 1.6 | 1.6 |
| **Distraction** | **% time the head was oriented toward smartphone during waiting time** *(%)* | | Smartphone | 72.9 | 68.0 | 76.2 |
| | | | Distracted & safety measure | 74.7 | 69.8 | 78.0 |
| | **% time the head was oriented toward smartphone during crossing** *(%)* | | Smartphone | 73.5 | 73.1 | 73.7 |
| | | | Distracted & safety measure | 69.6 | 71.5 | 68.3 |
| | **# of head orientations to smartphone during a trial** *(N/s)* | | Smartphone | 0.2 | 0.2 | 0.2 |
| | | | Distracted & safety measure | 0.2 | 0.2 | 0.1 |
| **Safety Measures** | **Maximum acceleration** *(m/s$^2$)* | | Not distracted | 6.1 | 4.8 | 6.9 |
| | | | Smartphone | 5.1 | 4.5 | 5.5 |
| | | | Distracted & safety measure | 5.2 | 5.5 | 5.0 |
| | **Maximum deceleration** *(m/s$^2$)* | | Not distracted | 8.5 | 4.9 | 10.9 |
| | | | Smartphone | 12.8 | 4.3 | 18.6 |
| | | | Distracted & safety measure | 5.0 | 4.6 | 5.3 |
| | **Trial crossing label** | **Successful** *(%)* | Not distracted | 76.9 | 70.6 | 81.2 |
| | | | Smartphone | 61.4 | 64.7 | 59.2 |
| | | | Distracted & safety measure | 68.8 | 65.9 | 70.8 |
| | | **Time-out** *(%)* | Not distracted | 0.5 | 1.2 | 0.0 |
| | | | Smartphone | 1.9 | 3.5 | 0.8 |
| | | | Distracted & safety measure | 1.0 | 1.8 | 0.4 |
| | | **Failed** *(%)* | Not distracted | 22.6 | 28.2 | 18.8 |
| | | | Smartphone | 36.9 | 31.8 | 40.4 |
| | | | Distracted & safety measure | 30.2 | 32.4 | 28.8 |
| | **Minimum PET** *(S)* | | Not distracted | 1.2 | 1.1 | 1.2 |
| | | | Smartphone | 1.0 | 1.1 | 1.0 |
| | | | Distracted & safety measure | 1.1 | 1.2 | 1.1 |

**Table 1: Pedestrian Behavioural Attributes across Three Crossing Conditions**

Comparing the success, failure and time-out rates proves some interesting results. It is observed (Table 1), as expected, non-distracted participants had a significantly higher success rate compared to their distracted conditions which is 18% higher for males compared to females. The success rate in condition three (the smart LED treatment) improved by 7.4% compared to condition two which was seen mostly in the male participants which indicates that men responded more positively to the treatment.

Pedestrian accelerations and decelerations are one of the considered variables which represent the walking behaviour of participants. While there is a consistent decrease in acceleration rate across all conditions, the general trend shows that distracted participants had a higher maximum deceleration compared to the not distracted situation, while the lowest value for this variable was observed in the smart LED safety treatment condition. This shows that there is a high speed-variability among distracted pedestrians. Females have a lower maximum acceleration and deceleration compared to males. Average acceleration and deceleration is almost consistent through all conditions, while for each gender it is observed that males have slightly higher average acceleration compared to females.

The final considered behavioural variables are safety measures of PET and TTC. Throughout our analysis, there was no significant TTC. For the PET component, the lowest recorded time was 1 second which is considered a serious conflict, this observation was recorded in the second condition with the distracted pedestrians and no safety treatment. Moreover, the smart LED treatment increased the value of this safety measure by 13% compared to the untreated condition.

To better understand and analyse the effect of the smartphone distraction and the proposed safety encounter on the sample, we generated graphs presented in Figure 6 which illustrate meaningful interactions between less severe PETs (PET of more than 1.5 seconds) and some of the study variables.

4.2.1 <u>Safety Crossing Pattern Sensitivity Analysis</u>

Figure 6.a shows that females' safety decreases significantly when they are distracted; while, the smart LED light improved their crossing safety (2% increase). A similar pattern can be seen for men. Although comparing the gender's pattern, we observe that males in general crossed more carefully, while the distraction had a more negative impact on them compared to women (28% versus 12% decrease). An explanation to this fact can be that females are better in multitasking compared to the opposite gender.

Interactions between PET less than 1.5 seconds and the age factor indicates that people who are more than 30 years old were less successful in walking across the street in a safely manner compared to younger adults (Figure 6.b). However, the LED light smoothened the negative impact of the distraction compared to the untreated condition.

Investigating the crossing pattern of the pedestrians who waited on the sidewalk more than average (i.e. more than 20 seconds), confirms that although the safety treatment increased their safety compared to condition two, their possibility of passing the road was reduced considerably compared to when they were not distracted (Figure 1.c).

As presented in Figure 6.d and 6.e, when pedestrians hassled to start or finish their crossing, they potentially were risking their safety, especially when they were distracted. Interestingly, the smart LED light helped the distracted individuals who were in a rush to cross (Figure 6.d).

Looking at the effects of smartphone distraction on pedestrians in Figure 6.f to 6.h it is observed that although the proposed treatment enhanced the crossing performance, interestingly it did not change the crossing behaviour of very discarded individuals (i.e. people who spent more than 75% of their time on their phone either while waiting on the sidewalk or crossing).

In summary, descriptive and sensitivity analysis of the crossing tracking information across all three conditions, shows that indeed distracted pedestrians have lower PET measures despite the fact that as discussed above, participants had higher speeds in the distracted conditions proving that distraction has a negative effect on safety which supports past findings (Byington & Schwebel, 2013; M. B. Neider et al., 2010; Schwebel et al., 2012). Also, evidently, the proposed safety measure improved the PET measure compared to the distracted condition

To better capture the real effect of the study variables on pedestrian crossing behaviour, the model results regarding the effect of variables on PET are presented in the in the next section.

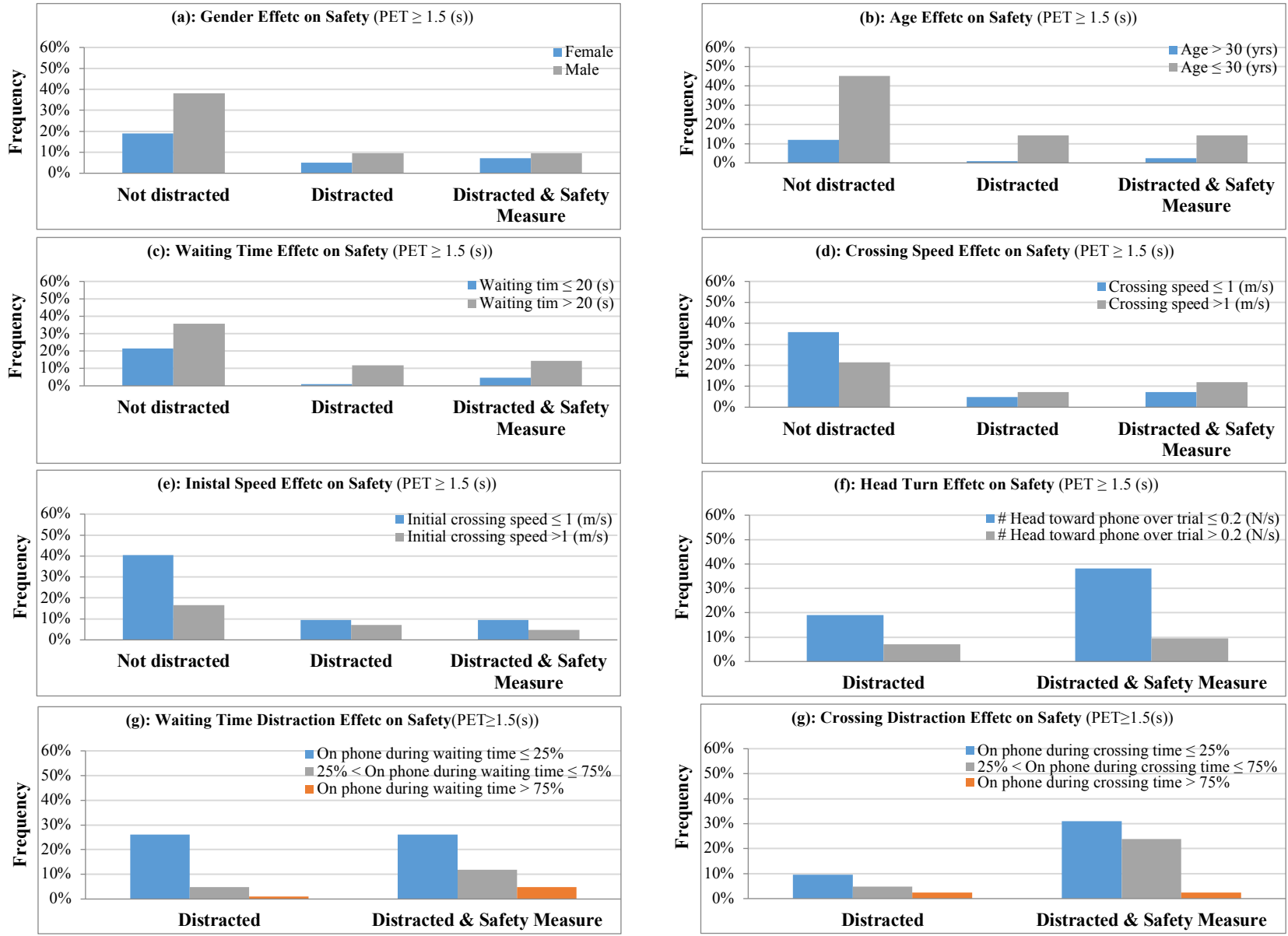

Figure 6: Sample Sensitivity Analysis

## 5 MODEL RESULTS AND DISCUSSION

### 5.1 Model Parameters

To estimate the effects of variables on the pedestrians' crossing safety, we employed the MNL model (Table 2). The dependent variable was computed as the safe event when the minimum recorded PET was more than 1.5 seconds. The model fit measures show that all three MNL models fit our dataset and the base condition with no distraction fit our data best.

As evident from Table 2, females were more careless compared to men. Comparing their behaviour pattern across the three conditions indicates that distraction has a negative effect on their crossing while smart LED light treatments reduced the negative impact compared to the no treatment condition. As expected, waiting time duration has a positive impact on pedestrian safety. This is due to the fact that while they are on the sidewalk, they can observe and register what is happening in the environment around them and make a safer crossing decision. In the smartphone distraction condition wait time had a lower safety improvement effect compared to the not distracted and smart LED implemented condition.

Three speed variables that have significant impact on the crossing behaviour of pedestrians are the initial walking speed, average deceleration and average acceleration. The first two attributes decreased the propensity of crossing safely, unlike the average acceleration factor. This can be explained since individuals want to cross the dangerous zone faster and while they are distracted their sudden movements might lead to unsafe crossing. The value of the coefficients across the three conditions provides interesting findings where the proposed treatment smoothens the negative impact of these attributes.

Number of head turns during the trial has a positive impact on crossing safely which means while pedestrians were distracted by their phone, they were looking back and forth at the road to observe the ongoing traffic for a safe gap. And finally, as expected, the percentage of the time spent looking at the smartphone and away from the road while standing at the cross walk increased the probability of unsafe crossing. That being said, the smart LED treatment improved this negative impact.

At the end by comparing the estimated parameters values across the three models, it can be observed that across the three crossing conditions, the smart LED light improved the crossing performance compared to the untreated condition.

| Crossing conditions<br>Variables | Not distracted | Distracted | Distracted & Safety Measure |
|---|---|---|---|
| **Female** | -0.11 (-1.88) | -0.21 (-1.76) | -0.10 (-1.52) |
| **Wait time duration** | 0.02 (3.34) | 0.01 (1.83) | 0.02 (2.67) |
| **Initial walking speed** | -0.65 (-1.85) | -2.11 (-2.14) | -0.49 (-1.47) |
| **Average acceleration** | 1.25 (1.87) | 1.75 (1.72) | 1.14 (1.98) |
| **Average deceleration** | -0.19 (-1.76) | -0.98 (-2.07) | -0.63 (-1.62) |
| **Head turn during a trial** | - | 2.53 (1.53) | 2.82 (1.37) |
| **# of head orientations to smartphone during a trial** | - | -5.33 (-1.42) | -1.54 (-1.52) |
| **% time the head was oriented toward smartphone during waiting time** | - | -6.67 (-1.50) | -4.47 (-2.82) |
| $\rho^2$ | 0.18 | 0.13 | 0.17 |

Table 2: Effects of Exogenous and Distraction Attributes on Pedestrian Crossing across Three Crossing Conditions

## 6 CONCLUSION

Increasing pedestrian fatalities have been linked to the increase in distracted pedestrian and drivers on the road. Studies to further investigate this trend as well as the effectiveness of countermeasures are of a risky nature given the close interaction of vulnerable road users with motorized vehicle. For this reason, we employ the use of a Head-Mounted Immersive Virtual Reality (VIRE) to simulate an existing road crossing in Montréal with real-time traffic flows. This provides a safe environment for performing any type of controlled safety study. Since past studies have identified looking down at one's phone as the riskiest form of distraction, our study made use of a maze-solving puzzle in on smartphone as the distraction source.

Most preventative measure in this field are implemented as mobile applications to alarm the distracted pedestrian when crossing the road. The limitation with these approaches is that although they are available to download, most individuals will not install them. To overcome this constraint, our proposed treatment is to install smart flashing and color changing LED lights at road crossings to alert pedestrian who are distracted by their phones by means of flashing lights which is visible to pedestrian whether they are looking at their smartphone, talking on their phones or listening to music. Our methodology provides a useful tool to evaluate street crossing behaviour and safety in a safe and controlled environment where many different scenarios, safety treatments and crossing movements can be tested.

In our study, participants performed three conditions of road crossing tasks while 1) control condition, crossing the road with no distraction 2) distracted condition, solving a maze-puzzle on a smartphone and 3) distracted condition with implemented safety measure: crossing the road while solving a maze-puzzle on a smartphone with implemented smart LED lights on the crosswalk to warn them of unsafe crossing initiations. The descriptive analysis of the crossing shows that indeed in the distracted condition, pedestrians have lower PET measures and higher speeds, suggesting that distraction has a negative effect on safety. The proposed safety measure condition improved the PET measures compared to the distracted condition, establishing its effectiveness in preventing distracted pedestrians from initiating dangerous crossings.

A brief overview of the MNL model results indicates that females have more dangerous crossings especially in distracted conditions; however, the smart LED treatment reduces the negative impact. Average deceleration and initial walking speed have a negative impact on pedestrian crossing behaviour. The distraction factor significantly increases the negative impact while the smart LED treatment improves the effects of these variables on safe crossing. Studying the impacts of distraction attributes on the safe crossing of distracted pedestrians demonstrates that the number of times the pedestrian's head is facing the smartphone during a trial and percentage of time the head was facing the smartphone during waiting time increases the possibility of unsafe crossings, however, the proposed treatment has less severe impacts. Overall, both descriptive analysis and model results proves the efficiency of our proposed safety measure of colour changing smart LED lights on the crosswalk on improving the safety of distracted pedestrian crossings.

Our study is not without limitation which can be addressed in the future and extended studies. In the simulation system, the interaction between pedestrian and mixed traffic (vehicles, pedestrians, and cyclists) can be studied. Also, the eye traction can be added as a measure to generate more distraction attributes, and the distraction effects of the smart LED lights should be further studied. The comparison between different types of distraction on pedestrian's crossing behaviour such as listening to music, calling, texting etc. can also be examined in the VIRE. Finally, with the use of this methodological approach, many new safety treatments can be implemented and evaluated in the VIRE.

## ACKNOWLEDGMENT

The authors would like to acknowledge the financial support from the NSERC Discovery grant. Special thanks to Zihui Zhong and Alexandra Beaulieu for helping with conducting the experiments as well as Matin Nabavi Niaki and Sohail Zangenehpour for their constructive advice.

**REFERENCES**

Anciaes, P. R., & Jones, P. (2018). Estimating preferences for different types of pedestrian crossing facilities. *Transportation Research Part F: Traffic Psychology and Behaviour*, *52*, 222–237. https://doi.org/10.1016/j.trf.2017.11.025

Animesh, A., Pinsonneault, A., Yang, S. B., & Oh, W. (2011). An odyssey into virtual worlds: Exploring the impacts of technological and spatial environments on intention to purchase virtual products. *MIS Quarterly: Management Information Systems*, *35*(3), 789–810. https://doi.org/Article

Banducci, S. E., Ward, N., Gaspar, J. G., Schab, K. R., Crowell, J. A., Kaczmarski, H., & Kramer, A. F. (2016). The effects of cell phone and text message conversations on simulated street crossing. *Human Factors*, *58*(1), 150–162. https://doi.org/10.1177/0018720815609501

Blissing, B., Bruzelius, F., & Eriksson, O. (2017). Driver behavior in mixed and virtual reality - A comparative study. *Transportation Research Part F: Traffic Psychology and Behaviour*. https://doi.org/10.1016/j.trf.2017.08.005

Brisan, C., Vasiu, R. V., & Munteanu, L. (2013). A modular road auto-generating algorithm for developing the road models for driving simulators. *Transportation Research Part C: Emerging Technologies*, *26*, 269–284. https://doi.org/10.1016/j.trc.2012.09.008

Byington, K. W., & Schwebel, D. C. (2013). Effects of mobile Internet use on college student pedestrian injury risk. *Accident Analysis and Prevention*, *51*, 78–83. https://doi.org/10.1016/j.aap.2012.11.001

Chen, X., Zhu, Y., & Wang, G. (2012). Evaluating a mobile pedestrian safety application in a virtual urban environment. *Proceedings of the 11th ACM SIGGRAPH International Conference on Virtual-Reality Continuum and Its Applications in Industry - VRCAI '12*, *1*(212), 175–180. https://doi.org/10.1145/2407516.2407561

Cunningham, M. ., & Regan, M. (2018). Driver Distraction and Inattention. In *Safe Mobility: Challenges, Methodology and Solutions* (pp. 57–82). Emerald Publishing Limited.

Cunningham, M., & Regan, M. (2017). Driver Distraction and Inattention in the Realm of Automated Driving. *IET Intelligent Transport Systems*, (January). https://doi.org/10.1049/iet-its.2017.0232

Dommes, A., Cavallo, V., Dubuisson, J.-B., Tournier, I., & Vienne, F. (2014). Crossing a two-way street: comparison of young and old pedestrians. *Journal of Safety Research*, *50*, 27–34.

Faiola, A., Newlon, C., Pfaff, M., & Smyslova, O. (2012). Correlating the effects of flow and telepresence in virtual world: enhancing our understanding of user behavior in game-based learning. *Computers in Human B*, *29*, 1113–1121. https://doi.org/10.1016/j.chb.2012.10.003

Farooq, B., Cherchi, E., & Sobhani, A. (2018). Virtual Immersive Reality for Stated Preference Travel Behaviour Experiments: A Case Study of Autonomous Vehicles on Urban Roads. *Transportation Research Record*. Retrieved from https://arxiv.org/ftp/arxiv/papers/1802/1802.06180.pdf

Feldstein, I., Lehsing, C., Dietrich, A., & Bengler, K. (2016). Pedestrian Simulators for Traffic Research : State of the Art and Future of a Motion Lab. In *4th International Digital Human Modeling Symposium*.

Gaspar, J. G., Neider, M. B., Crowell, J. A., Lutz, A., Kaczmarski, H., & Kramer, A. F. (2013). Are gamers better crossers? An examination of action video game experience and dual task effects in a simulated street crossing task. *Human Factors: The Journal of the Human Factors and Ergonomics Society*, *56*, 443–452. https://doi.org/10.1177/0018720813499930

Gitelman, V., Carmel, R., Pesahov, F., & Hakkert, S. (2017). An examination of the influence of crosswalk marking removal on pedestrian safety as reflected in road user behaviours. *Transportation Research Part F: Traffic Psychology and Behaviour*, *46*, 342–355. https://doi.org/10.1016/j.trf.2016.03.007

Hacohen, S., Shvalb, N., & Shoval, S. (2018). Dynamic model for pedestrian crossing in congested traffic based on probabilistic navigation function. *Transportation Research Part C: Emerging Technologies*, *86*(March 2017), 78–96. https://doi.org/10.1016/j.trc.2017.10.024

Hakkert, A. S., Gitelman, V., & Ben-Shabat, E. (2002). An evaluation of crosswalk warning systems: Effects on pedestrian and vehicle behaviour. *Transportation Research Part F: Traffic Psychology and Behaviour*, *5*(4), 275–292. https://doi.org/10.1016/S1369-8478(02)00033-5

Haque, M. M., & Washington, S. (2015). The impact of mobile phone distraction on the braking behaviour of young drivers: A hazard-based duration model. *Transportation Research Part C: Emerging Technologies*, *50*, 13–27. https://doi.org/10.1016/j.trc.2014.07.011

Hatfield, J., & Murphy, S. (2007). The effects of mobile phone use on pedestrian crossing behaviour at signalised and unsignalised intersections. *Accident Analysis and Prevention*, *39*, 197–205. https://doi.org/10.1016/j.aap.2006.07.001

Holland, C., & Hill, R. (2010). Gender differences in factors predicting unsafe crossing decisions in adult pedestrians across the lifespan: A simulation study. *Accident Analysis & Prevention*, *42*, 1097–1106.

Hwang, T., Jeong, J. P., & Lee, E. (2014). SANA: Safety-Aware Navigation App for pedestrian protection in vehicular networks. *International Conference on ICT Convergence*. https://doi.org/10.1109/ICTC.2014.6983341

Ifedi, F. (2005). *Sport participation in Canada, 2005*. Statistics Canada: Culture, Tourism and Centre for Education Statistics Division.

Jennett, C., Cox, A. L., Cairns, P., Dhoparee, S., Epps, A., Tijs, T., & Walton, A. (2008). Measuring and defining the experience of immersion in games. *International Journal of Human Computer Studies*, *66*(9), 641–661. https://doi.org/10.1016/j.ijhcs.2008.04.004

Knoblauch, R. L., Pietrucha, M. T., & Nitzburg, M. (1996). Field studies of pedestrian walking speed and start-up time. *Transportation Research Record: Journal of the Transportation Research Board*, *1538*, 1996.

Kramer, A., & Madden, D. (2008). *The handbook of aging and cognition* (3rd ed.). Psychology Press.

Lehsing, C., & Feldstein, I. T. (2018). Urban Interaction – Getting Vulnerable Road Users into Driving Simulation. In *UR:BAN Human Factors in Traffic* (pp. 347–362). https://doi.org/10.1007/978-3-658-15418-9_19

Lin, M.-I. B., & Huang, Y.-P. (2017). The impact of walking while using a smartphone on pedestrians' awareness of roadside events. *Accident Analysis & Prevention*, *101*, 87–96. https://doi.org/10.1016/j.aap.2017.02.005

Mcdowell, A., & Shi, A. (2014). *Introducing the BCHOICE Procedure for Bayesian Discrete Choice Models*.

Muehlbacher, D., Preuk, K., Lehsing, C., Will, S., & Dotzauer, M. (2018). Multi-Road User Simulation: Methodological Considerations from Study Planning to Data Analysis. In *UR:BAN Human Factors in Traffic* (pp. 403–418). https://doi.org/https://doi.org/10.1007/978-3-658-15418-9_23

Nah, F. F.-H., Eschenbrenner, B., & DeWester, D. (2011). Enhancing brand equity through flow and telepresence: A comparison of 2D and 3D virtual worlds. *MIS Quarterly*, *35*(3), 1–19. https://doi.org/10.1002/mrdd.20052

Nasar, J., Hecht, P., & Wener, R. (2008). Mobile telephones, distracted attention, and pedestrian safety. *Accident Analysis and Prevention*, *40*(1), 69–75. https://doi.org/10.1016/j.aap.2007.04.005

Nasar, J. L., & Troyer, D. (2013). Pedestrian injuries due to mobile phone use in public places. *Accident Analysis and Prevention*, *57*, 91–95. https://doi.org/10.1016/j.aap.2013.03.021

Neider, M. B., Gaspar, J. G., Mccarley, J. S., Crowell, J. A., Kaczmarski, H., & Kramer, A. F. (2011). Walking & talking: dual-task effects on street crossing behavior in older adults. *Psychol Aging*, *26*(2), 260–268. https://doi.org/10.1037/a0021566.Walking

Neider, M. B., McCarley, J. S., Crowell, J. A., Kaczmarski, H., & Kramer, A. F. (2010). Pedestrians, vehicles, and cell phones. *Accident Analysis and Prevention*, *42*, 589–594. https://doi.org/10.1016/j.

O'Neill, M. J. (1992). Effects of familiarity and plan complexity on wayfinding in simulated buildings. *Journal of Environmental Psychology*, *12*(4), 319–327. https://doi.org/10.1016/S0272-4944(05)80080-5

Oculus Gear VR. (n.d.). Retrieved December 13, 2017, from https://www.oculus.com/gear-vr/

Olshannikova, E., Ometov, A., Koucheryavy, Y., & Olsson, T. (2015). Visualizing Big Data with augmented and virtual reality: challenges and research agenda. *Journal of Big Data*, *2*(1), 1–27. https://doi.org/10.1186/s40537-015-0031-2

Oviedo-Trespalacios, O., Haque, M. M., King, M., & Washington, S. (2016). Understanding the impacts of mobile phone distraction on driving performance: A systematic review. *Transportation Research Part C: Emerging Technologies*, *72*, 360–380. https://doi.org/10.1016/j.trc.2016.10.006

Patterson, Z., Darbani, J. M., Rezaei, A., Zacharias, J., & Yazdizadeh, A. (2017). Comparing text-only and virtual reality discrete choice experiments of neighbourhood choice. *Landscape and Urban Planning*, *157*, 63–74. https://doi.org/10.1016/j.landurbplan.2016.05.024

Perroud, B., Régnier, S., Kemeny, A., & Mérienne, F. (2017). Model of realism score for immersive VR systems. *Transportation Research Part F: Traffic Psychology and Behaviour*. https://doi.org/10.1016/j.trf.2017.08.015

Plumert, J. M., & Kearney, J. K. (2014). Linking Decisions and Actions in Dynamic Environments: How Child and Adult Cyclists Cross Roads With Traffic. *Ecological Psychology*, *26*(September), 125–133.

https://doi.org/10.1080/10407413.2014.874933

Ruddle, R. A., Payne, S. J., & Jones, D. M. (1997). Navigating buildings in "dek-top" virtual environments: Experimental investigations using extended navigational experience. *Journal of Experimental Psychology: Applied*, *3*(2), 143–159. https://doi.org/10.1037/1076-898X.3.2.143

Rusch, M. L., Schall, M. C., Gavin, P., Lee, J. D., Dawson, J. D., Vecera, S., & Rizzo, M. (2013). Directing driver attention with augmented reality cues. *Transportation Research Part F: Traffic Psychology and Behaviour*, *16*, 127–137. https://doi.org/10.1016/j.trf.2012.08.007

Schwebel, D. C., Stavrinos, D., Byington, K. W., Davis, T., O'Neal, E. E., & De Jong, D. (2012). Distraction and pedestrian safety: How talking on the phone, texting, and listening to music impact crossing the street. *Accident Analysis and Prevention*, *45*, 266–271. https://doi.org/10.1016/j.aap.2011.07.011

Sobhani, A., Farooq, B., & Zhong, Z. (2017). Immersive head mounted virtual reality based safety analysis of smartphone distracted pedestrians at street crossing. In *Road Safety and simulation International Conference*. The Hague, Netherlands.

Song, Y. E., Lehsing, C., Fuest, T., & Bengler, K. (2018). External HMIs and their effect on the interaction between pedestrians and automated vehicles. In *Advances in Intelligent Systems and Computing* (Vol. 722, pp. 13–18). https://doi.org/10.1007/978-3-319-73888-8_3

Stavrinos, D., Byington, K. W., & Schwebel, D. C. (2009). Effect of cell phone distraction on pediatric pedestrian injury risk. *Pediatrics*, *123*(2), 179–85. https://doi.org/10.1542/peds.2008-1382

Stavrinos, D., Byington, K. W., & Schwebel, D. C. (2011). Distracted walking: cell phones increase injury risk for college pedestrians. *Journal of Safety Research*, *42*(2), 101–107. https://doi.org/10.1016/j.jsr.2011.01.004

Tapiro, H., Oron-Gilad, T., & Parmet, Y. (2016). Cell phone conversations and child pedestrian's crossing behavior; a simulator study. *Safety Science*, *89*, 36–44. https://doi.org/10.1016/j.ssci.2016.05.013

Tlauka, M., & Wilson, P. N. (1996). Orientation-free representations from navigation through a computer-simulated environment. *Environment and Behaviour*, *28*(5), 647–664. https://doi.org/0803973233

Transport Canada. (2010). Un bref aperçu des usagers de la route vulnérables qui sont victimes d'accidents mortels.

Transport Canada. (2015). Canadian Motor Vehicle Traffic Collision Statistics: 2015.

Wang, Q., Guo, B., Peng, G., Zhou, G., & Yu, Z. (2016). CrowdWatch: Pedestrian safety assistance with mobile crowd sensing. *UbiComp 2016 Adjunct - Proceedings of the 2016 ACM International Joint Conference on Pervasive and Ubiquitous Computing*, 217–220. https://doi.org/10.1145/2968219.2971433

Wang, T., Cardone, G., Corradi, A., Torresani, L., & Campbell, A. T. (2012). WalkSafe: A Pedestrian Safety App for Mobile Phone Users WhoWalk and Talk While Crossing Roads. *Proceedings of the Twelfth Workshop on Mobile Computing Systems & Applications - HotMobile '12*, 1. https://doi.org/10.1145/2162081.2162089

Wang, Y., Liu, W., Meng, X., Fu, H., Zhang, D., Kang, Y., … Jiang, G. (2016). Development of an immersive virtual reality head-mounted display with high performance. *Applied Optics*, *55*(25), 6969–6977. https://doi.org/10.1364/AO.55.006969

Wiener, J., Hölscher, C., Büchner, S., & Konieczny, L. (2012). Gaze behavior during space perception and spatial decision making. *Psychological Research*, *76*(6), 713–729.

World Health Organization. (2016). Road Traffic Injuries.

Wu, H., Ashmead, D. h, & Bodenheimer, B. (2009). Using immersive virtual reality to evaluate pedestrian street crossing decisions at a roundabout. In *6th Symposium on Applied Perception in Graphics and Visualization*.

Yager, C. E., Cooper, J. M., & Chrysler, S. T. (2012). The Effects of Reading and Writing Text-Based Messages While Driving. *Proceedings of the Human Factors and Ergonomics Society Annual Meeting*, *56*(1), 2196–2200.

Ye, H., Xiao, F., & Yang, H. (2017). Exploration of day-to-day route choice models by a virtual experiment. *Transportation Research Procedia*, *23*, 679–699. https://doi.org/10.1016/j.trpro.2017.05.038

Zaki, M. H., & Sayed, T. (2016). Exploring walking gait features for the automated recognition of distracted pedestrians. *IET Intelligent Transport Systems*, *10*(2), 106–113. https://doi.org/10.1049/iet-its.2015.0001

Zhou, Z. (2015). HeadsUp: Keeping pedestrian phone addicts from dangers using mobile phone sensors. *International Journal of Distributed Sensor Networks*, *2015*, 1–9. https://doi.org/10.1155/2015/279846